\documentclass{aa} 
\usepackage{txfonts}
\usepackage{natbib,graphicx,amssymb}
\usepackage[english]{babel}
\def\sagelong{\object{SAGE1CJ053634.78-722658.5}}
\def\sage05{SAGE1CJ053634}

\begin{document}

\title{The Spitzer discovery of a galaxy with infrared emission solely
  due to AGN activity. \thanks{based on observations obtained with
    Spitzer.}}

\author{
S.~Hony\inst{1},
F.~Kemper\inst{2,3},
Paul~M.~Woods\inst{3},
J.~Th.~van Loon\inst{4},
V.~Gorjian\inst{5},
S.~C.~Madden\inst{1},
A.~A.~Zijlstra\inst{3},
K.~D.~Gordon\inst{6},
R.~Indebetouw\inst{7},
M.~Marengo\inst{8},
M.~Meixner\inst{6},
P.~Panuzzo\inst{1},
B.~Shiao\inst{6},
G.~C.~Sloan\inst{9},
J.~Roman-Duval\inst{6},
J.~Mullaney\inst{1},
\and
A.~G.~G.~M.~Tielens\inst{10}
  }
  
  \authorrunning{Hony et al.}
  \titlerunning{An AGN dominated galaxy}
  \offprints{S. Hony (sacha.hony@cea.fr)} 
  
\institute{
Laboratoire AIM Paris-Saclay, CNRS/INSU CEA/Irfu Université Paris
Diderot, B\^{a}t. 709, Orme des Merisiers, 91191, Gif sur Yvette
C{\'e}dex, France \email{sacha.hony@cea.fr}
\and
Academia Sinica Institute of Astronomy and Astrophysics, P.O. Box
23-141, Taipei 10617, Taiwan, R.O.C.
\and
Jodrell Bank Centre for Astrophysics, Alan Turing
  Building, School of Physics and Astronomy, University of
  Manchester, Oxford Road, Manchester, M13 9PL, UK
\and
Astrophysics Group, Lennard-Jones Laboratories, Keele
University, Staffordshire ST5 5BG, UK
\and
MS 169-506, Jet Propulsion Laboratory, California
  Institute of Technology, 4800 Oak Grove Drive, Pasadena, CA 91109
\and
Space Telescope Science Institute, 3700 San Martin
  Drive, Baltimore, MD 21218
\and
Department of Astronomy, P.O. Box 400325, University
  of Virginia, Charlottesville, VA 22904
\and
Department of Physics and Astronomy, Iowa State University, A313E
Zaffarano, Ames, IA 50010
\and
Center for Radiophysics and Space Research, Cornell
  University, Ithaca, NY 14853
\and
Leiden Observatory, P.O. Box 9513, 2300 RA Leiden, The Netherlands
}
  
  \date{\today;\today}
  % \abstract{}{}{}{}{} 
  % 5 {} token are mandatory
  
  \abstract
  {}
  {We present an analysis of a galaxy (\sagelong) at a redshift of
    0.14 of which the infrared (IR) emission is entirely dominated
    by emission associated with the active galactic nucleus.}
  {We present the 5$-$37~$\mu$m {\it Spitzer}/IRS spectrum and broad
    wavelength spectral energy distribution (SED) of \sagelong, an IR
    point-source detected by {\it Spitzer}/SAGE (Meixner et al 2006).
    The source was observed in the SAGE-Spec program (Kemper et al.,
    2010) and was included to determine the nature of sources with
    deviant IR colours. The spectrum shows a redshifted ($z$=0.14$\pm$0.005)
    silicate emission feature with an exceptionally high
    feature-to-continuum ratio and weak polycyclic aromatic
    hydrocarbon (PAH) emission bands. We compare the source with
    models of emission from dusty tori around AGNs from Nenkova et al.
    (2008). We present a diagnostic diagram that will help to identify
    similar sources based on {\it Spitzer}/MIPS and {\it
      Herschel}/PACS photometry.}
  {The SED of \sagelong\ is peculiar because it lacks far-IR emission
    { due to cold dust} and a clear stellar counterpart. We find that
    the SED and the IR spectrum can be understood as emission
    originating from the inner $\sim$10~pc around an accreting black
    hole. There is no need to invoke emission from the host galaxy,
    either from the stars or from the interstellar medium, although a
    possible early-type host galaxy cannot be excluded based on the
    SED analysis. The hot dust around the accretion disk gives rise to
    a continuum, which peaks at 4~$\mu$m, whereas the strong silicate
    features may arise from optically thin emission of dusty clouds
    within $\sim$10~pc around the black hole. The weak PAH emission
    does not appear to be linked to star formation, as star formation
    templates strongly over-predict the measured far-IR flux levels. }
  {The SED of \sagelong\ is rare in the local universe but may be more
    common in the more distant universe. The conspicuous absence of
    host-galaxy IR emission places limits on the far-IR emission
    arising from the dusty torus alone.}

  
  \maketitle
  
\section{Introduction}
\label{sec:intro}
\begin{figure*}[!t]
  \includegraphics[width=\textwidth]{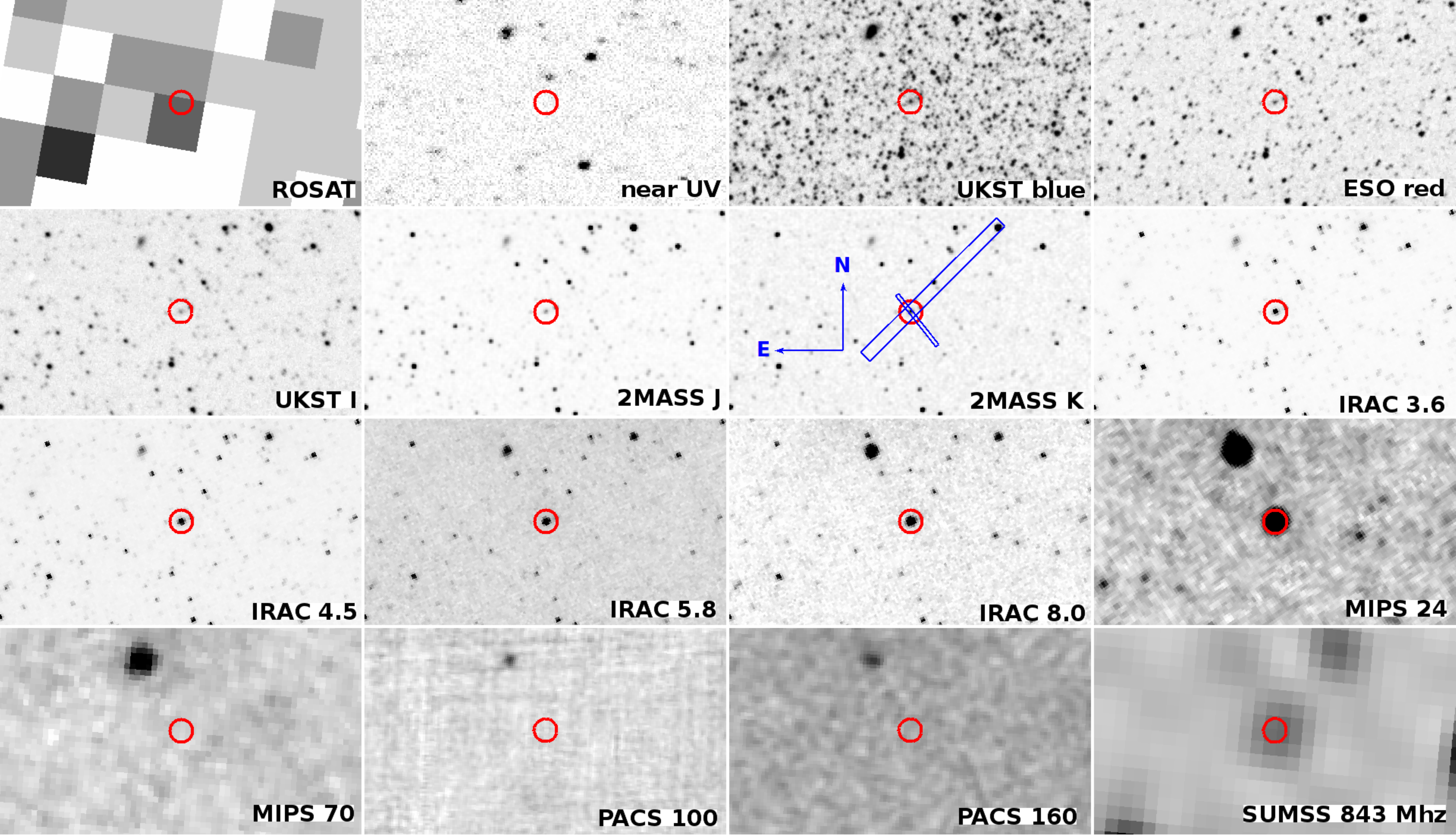}
  \caption{Catalogue of photometric images in increasing wavelength
    from the X-ray to radio, top left to bottom right. The position of
    \sage05 is marked by a small circle at the centre of each tile.
    The image orientation is indicated in the 2MASS K$_{\rm s}$ panel
    with arrows of length 1$^\prime$. The small and the large box in
    the 2MASS K$_{\rm s}$ panel indicate the slit position
    corresponding to the {\it Spitzer}/IRS short-low and long-low
    observations, respectively}
  \label{fig:maps}
\end{figure*}
The Spitzer SAGE survey of the Large Magellanic Cloud
\citep[LMC;][]{2006AJ....132.2268M}, hereafter referred to as SAGE-LMC
has provided a deep infrared (IR) photometric survey of $\sim$7$^{\rm
  o}$$\times$7$^{\rm o}$ covering the LMC in all {\it Spitzer}
\citep{2004ApJS..154....1W} Infrared Array Camera
\citep[IRAC;][]{2004ApJS..154...10F} and Multi-Band Imaging Photometer
for Spitzer \citep[MIPS;][]{2004ApJS..154...25R} photometric
wavelengths. Combined with existing Two Micron All Sky Survey
\citep[2MASS;][]{2006AJ....131.1163S}, the survey has yielded a
catalogue of $\sim$6,000,000 point-sources. Most of these are either
stars with an IR excess in the LMC, or foreground stars located in the
Milky Way. There is also a significant population of background
galaxies, uniformly distributed over the SAGE-LMC footprint
\citep{2008AJ....136...18W,2009ApJ...701..508K}. The SAGE-Spec program
\citep{2010PASP..122..683K} is a Spitzer spectroscopic follow-up
survey designed to verify the initial SAGE-LMC classifications
\citep{2006AJ....132.2268M,2006AJ....132.2034B,2008AJ....136...18W}
and investigate the nature of sources that could not be identified
based on their photometry. Here we report on a point-source, in the
SAGE-LMC
catalogue\footnote{http://irsa.ipac.caltech.edu/cgi-bin/Gator/nph-scan?submit=Select\&projshort=SPITZER}
identified as \sagelong\, (hereafter referred to as \sage05) at
moderate redshift ($z$=0.14) as derived from the observed wavelength
of the polycyclic aromatic hydrocarbon (PAH) features, which is
entirely dominated in its IR spectral energy distribution (SED) by
warm dust (silicate) emission, rather than emission from
star formation which is the case in nearly all other galaxies yet
observed. Here, we show that the infrared emission is arising solely
from dust heated by an active galactic nuclei (AGN), providing a
unique opportunity to study the AGN infrared SED without the
contaminating emission from the host galaxy.

\section{Results}
\label{sec:results}
\subsection{Observations}
\label{sec:observations}
The mid-IR spectrum of \sage05 ($\alpha$=05:36:34.358,
$\delta$=$-$72:26:57.50, J2000) is obtained using the InfraRed
Spectrograph \citep[IRS;][]{2004ApJS..154...18H} on board the Spitzer
Space Telescope \citep{2004ApJS..154....1W}. The observations and data
reduction were performed in the standard way for point-sources,
described by \citet{2010PASP..122..683K}. The position of the slit is
shown in the 2MASS-K$_{\rm s}$ panel of Fig.~\ref{fig:maps}.

\subsection{The mid-IR spectrum and SED}
\begin{figure}[!t]
  \includegraphics[width=8.8cm]{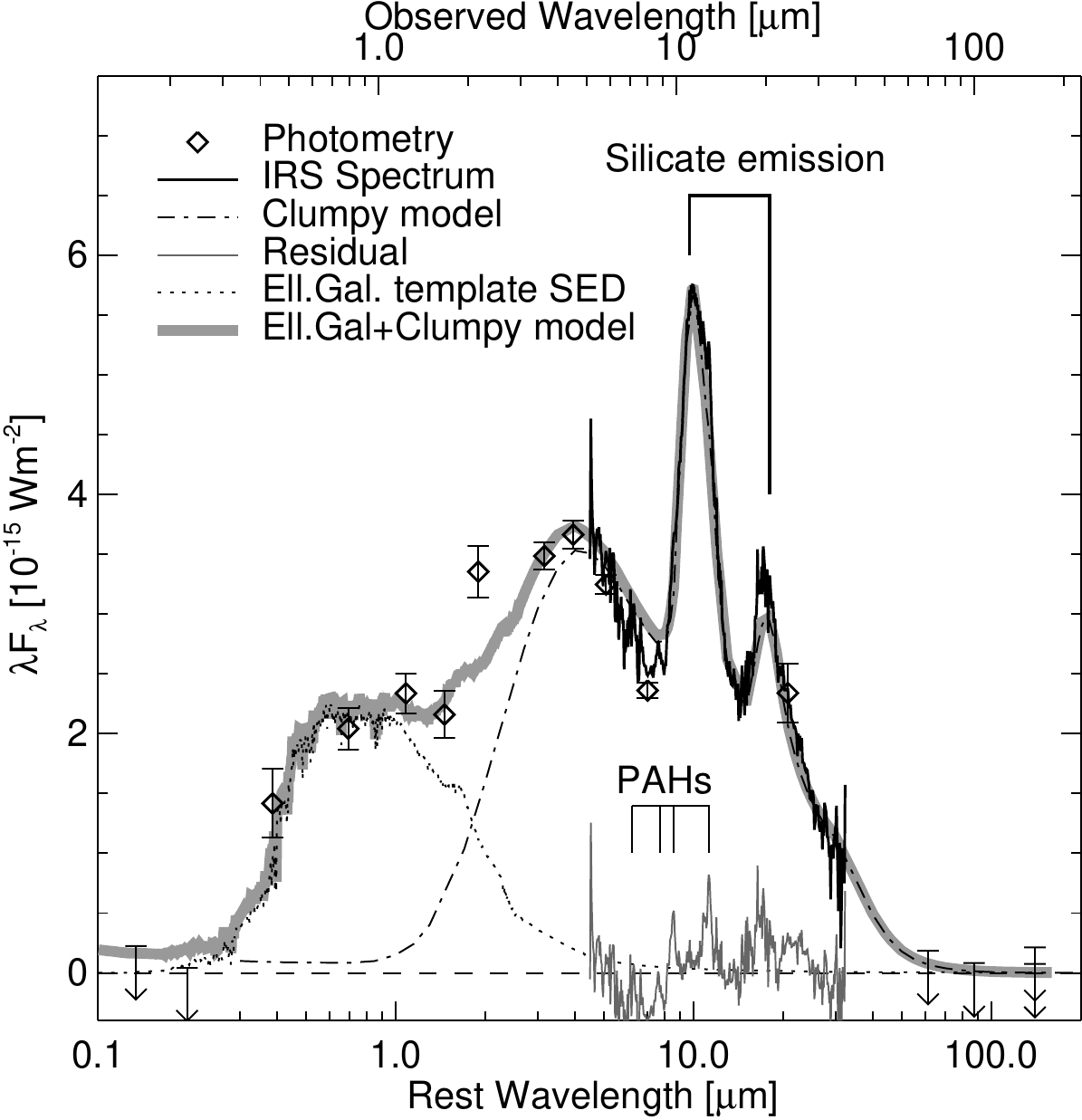}
  \caption{Energy distribution of \sage05. The diamonds indicate
    photometric measurements with their 1~$\sigma$ error-bars. The
    arrows indicate the 3~$\sigma$ upper limits. The predominance of
    warm silicate emission and the lack of cooler material in \sage05
    is evident. The dot-dashed line shows the best fitting clumpy
    torus model \citep{2008ApJ...685..160N}. The clumpy torus model
    reproduces the strength and band ratio of the silicate emission
    well. PAH emission can be identified, see also the residual (solid
    grey). The grey dotted line shows the possible contribution from
    an elliptical host galaxy. The thick grey line shows the sum of
    the clumpy torus model and host galaxy emission.}
  \label{fig:SED}
\end{figure}

The global SED and the IRS spectrum of \sage05 are presented in
Fig.~\ref{fig:SED}, shifted to rest-wavelength. The photometry
consists of IRAC and MIPS data from the SAGE-LMC project
\citep{2006AJ....132.2268M}, PACS data from the HERITAGE project
\citep{2010A&A...518L..71M} and near-IR data from the 2MASS project
\citep{2006AJ....131.1163S}, supplemented by near-UV
\citep[GALEX\footnote{{\tt
    http://galex.stsci.edu}};][]{2005ApJ...619L...1M}, optical
\citep[SuperCosmosSkySurvey\footnote{{\tt
    http://www-wfau.roe.ac.uk/sss}};][]{2000A&AS..144..235C,2008AJ....136..735L}
and a radio measurement
\citep{2003MNRAS.342.1117M,1995A&AS..111..311F}. Maps of the source
and its surroundings are shown in Fig.~\ref{fig:maps}.

The IRS spectrum (see Fig.~\ref{fig:SED}) shows weak PAH features at
the expected rest wavelengths of 6.23, 7.7, 8.6 and 11.25 $\mu$m, from
which a redshift of $z=0.14 \pm 0.005$ is derived. No convincing
detections of atomic transitions are present. The most striking
features in the IRS data are the very strong emission bands at 9.7 and
18~$\mu$m, due to the Si$-$O stretching and bending mode in silicates,
respectively. In addition, there is a weak 11.25~$\mu$m PAH emission
band at the long wavelength extreme of the 9.7~$\mu$m band. The SED
(Fig.~\ref{fig:SED}) is dominated by the mid-IR emission peaking
around 4~$\mu$m. There appears to be very little cooler material in
the galaxy as this would show up as emission at longer wavelengths
($>$50~$\mu$m). In fact, the flux of the IRS spectrum is already
declining at its longest wavelengths ($>$30~$\mu$m).

The radio flux densities are 40 and 13 mJy at 6.2 and 36 cm
\citep{1995A&AS..111..311F,2003MNRAS.342.1117M}, respectively. The
estimated $R$ value (ratio of 6~cm to 2500~\AA\ flux density) is
$\sim$200, making it a radio-loud source
\citep[e.g.][]{2007ApJ...656..680J}. The measurement at 6.2 cm is done
with a large beam which makes the association with this source
uncertain. The flux density at 36~cm, which is clearly coincident with
\sage05 (see Fig.~\ref{fig:maps}) still implies a radio-loud source.
There is also a possible coincidence with an X-ray source detected by
ROSAT \citep[][see Fig.~\ref{fig:maps}]{1999A&A...349..389V}. The main
observational characteristics of \sage05 are summarised in
Table~\ref{tab:characteristics}.

\subsection{Nature of the source}
\label{sec:nature-source}
\begin{figure*}[!t]
  \includegraphics[width=\textwidth]{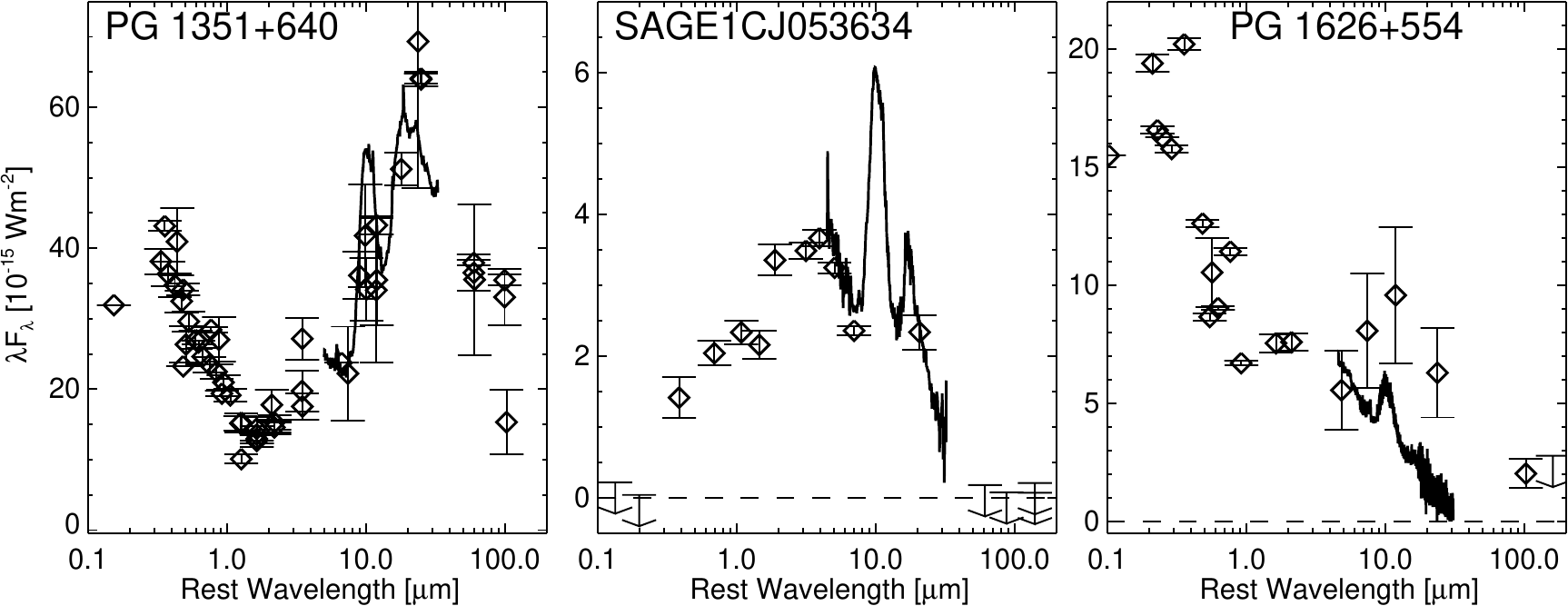}
  \caption{Comparison between \sage05 and the sources that exhibit the
    previously known extremes in either the 10~$\mu$m feature
    strength, \citep[PG 1351+640;][]{2005ApJ...625L..75H} or the
    decline towards the far-IR, \citep[PG
    1626+554;][]{2007ApJ...666..806N}. Both PG 1351+640 and PG
    1626+554 are peculiar sources, that are far from being
    proto-typical. The diamonds indicate photometric measurements with
    their 1~$\sigma$ error-bars and arrows indicate the 3~$\sigma$
    upper limits. The drawn lines show the IRS spectrum exhibiting the
    silicate emission features of each source. \sage05 stands out by
    the fact that most of its emission falls in the mid-IR range.}
  \label{fig:fig3}
\end{figure*}

The fact that the observed features in the spectrum are redshifted by
the amount observed implies that the source is not located in the
Milky Way or in the LMC. There are several indications that \sage05 is
a quasar: the presence of radio emission and the possible association
of an X-ray detection. Moreover, the IR SED exhibits the main
characteristics of other quasars. First, the luminosity of $\sim
10^{10}$ L$_{\sun}$, inferred from the X-ray and IR emission, is
within the typical range for quasars. Secondly, the emission bump
around 4~$\mu$m can be attributed to emission of the hottest dust in a
clumpy torus near the accretion disk around the central black hole
\citep[e.g.][]{2005AIPC..761..245S}. This emission feature is the
basis of the IRAC colour selection of active galaxies
\citep{2005ApJ...631..163S}. Finally, the silicate emission is a
characteristic of quasar sources
\citep[e.g.][]{2005A&A...436L...5S,2007ASPC..373..574H,2006ApJ...653..127S}
depending on the viewing angle of the source
\citep[e.g.][]{2006ApJ...653..127S}, albeit not exclusive to quasars,
as it is sometimes observed towards elliptical galaxies
\citep[e.g.][]{2006ApJ...639L..55B}. However, the ellipticals exhibit
significantly weaker silicate emission and the energy distribution is
dominated by the stellar emission of the evolved stars peaking around
1~$\mu$m (see also Fig.~\ref{fig:fig3}).

\subsection{Silicate emission}
\label{sec:silicate-emission}
\begin{table}[!t]
  \caption{Main observational characteristics of \sage05.}
  \begin{tabular}{l | l l l l }
    \hline
    Measurement & Value &Uncertainty & Unit & Remark \\
    \hline
    \hline
    redshift       &  0.14 & 0.005     &         & PAH bands      \\
    d$_{luminosity}$   & 654   & 25     & Mpc     &                \\
    $m$-$M$   & 39   &      & mag     &                \\
    $^{10}$log(L$_{X})$  &  9.6 &  & L$_{\sun}$     & ROSAT       \\
    $^{10}$log(L$_{FIR}$) & 10.0 &       & L$_{\sun}$ & 40-150~$\mu$m  \\
    $^{10}$log(L$_{IR}$) & 10.8 &       & L$_{\sun}$  & 8-1000~$\mu$m  \\
    far-UV         & $<$10 & (3$\sigma$)    & $\mu$Jy & GALEX         \\
    near-UV        & $<$3 & (3$\sigma$)   & $\mu$Jy & GALEX         \\
    B              & 18.514 &0.13 &  mag    & SuperCOSMOS      \\
    Gunn-i         & 16.819&0.092 &  mag    & DENIS         \\
    J              & 15.688&0.077 &  mag    & 2MASS         \\
    H              & 14.993&0.098 &  mag    & 2MASS         \\
    K$_{\rm s}$      & 13.743&0.070 &  mag    & 2MASS         \\
    IRAC 3.6       & 3.67&0.11    &  mJy    & Spitzer        \\
    IRAC 4.5       & 4.82&0.15    &  mJy    & Spitzer        \\
    IRAC 5.8       & 5.51&0.13    &  mJy    & Spitzer        \\
    IRAC 8.0       & 5.53&0.14    &  mJy    & Spitzer        \\
    MIPS 24        & 16.2&1.7     &  mJy    & Spitzer        \\
    MIPS 70        & $<$4 & (3$\sigma$)        &  mJy    & Spitzer        \\
    PACS 100      &  $<$3 & (3$\sigma$)        &   mJy    &  Herschel        \\
    MIPS 160       & $<$10 & (3$\sigma$)       &  mJy    & Spitzer        \\
    PACS 160       &  $<$4 & (3$\sigma$)       &   mJy    &  Herschel        \\
    4.75 GHz       & 40     &          &  mJy    & Parkes         \\
    % 2.45 GHz       & 102  &            &  mJy    & Parkes         \\
    843  MHz       & 13.8&1.3     &  mJy    & SUMSS          \\
    \hline
  \end{tabular}
  \label{tab:characteristics}
\end{table}
Two dust components can be distinguished in the spectrum. First, the
featureless emission component present in the SED, peaking at $\sim$4
$\mu$m (see Fig.~\ref{fig:SED}), is due to optically thick dust
emission from the accretion disk, corresponding to a dust temperature
of $\sim$1500~K. Second, the silicate emission bands at 10 and
18~$\mu$m due to emission of warm silicate grains.

We construct a very simple model to derive the main traits of the
observed energy distribution. The silicate emission feature is
modelled using optical properties of oxygen-rich glassy silicates
\citep[][ Set 2 in their Table~2 ]{1992A&A...261..567O}. We assume a
single temperature and optically thin emission which makes the
predicted spectrum the convolution of the absorption cross-section and
the Planck function of the corresponding temperature. The resulting
silicate emission is added to the optically thick continuum due to
warm dust in the accretion disk, represented in this case by a
constant 5.5~mJy spectrum. For a silicate temperature of 260~K --
consistent with the range found for other quasars
\citep[e.g.][]{2005ApJ...625L..75H} -- the simulated spectrum
reproduces the observed strengths of the 9.7 and 18 $\mu$m features.

The silicate emission is believed to arise from clouds of dust near
the central regions. The temperature of the silicate grains constrains
the distance at which these clouds are located. The models as
presented by \cite{2008ApJ...679..101S} shows that clouds around
5$-$10~pc away from the black hole show silicate temperatures as found
in the IRS spectrum of \sage05. \citet{2008ApJ...679..101S} have
modelled a sample of silicate emission quasars. Their results show that
the silicate feature-to-continuum ratio depends mostly on the cloud
distance to the black hole and less on the solid angle covered by
clouds.

We compare the observations with the CLUMPY models of
\citet{2008ApJ...685..160N}, which consist of an accreting black hole
surrounded by a torus that contains clouds of dust. We only show the
emission of the torus without the direct contribution from the AGN.
The observed emission from the AGN depends on the source inclination
angle for which we have no constraints. The calculated SEDs explain
the full IR emission of \sage05 well (see Fig.~\ref{fig:SED}) and we
need not invoke additional host galaxy emission to understand the
observed mid-IR to sub-mm SED. There is room for some emission from a
population of old stars in the rest frame optical range. The narrow
PAH bands can be identified more readily in the residual spectrum
obtained by subtracting the best fit model from the IRS spectrum.
There are a few broad structures in the residual, in particular the
model over-predicts the flux levels near 8~$\mu$m and under-predicts
the flux levels near 20~$\mu$m. These may reflect a slight mismatch in
the actual temperature distribution or silicate composition compared
to what is assumed in the CLUMPY model. The K$_{\rm s}$ band
photometry is not well reproduced which may also indicate that the
warmest dust component shines brighter than what is assumed in the
model.

The \citet{2008ApJ...685..160N} model contains many free parameters,
in particular those concerning the geometrical setup of the system,
that are ill-constrained. However, there are a few parameters that can
be determined. The main conclusions that can be drawn from the
comparison with the CLUMPY models are that the optical depth through
each individual cloud has to be moderate ($\tau_V \lesssim 15$), that
the clouds have to be located within a narrow distance range around
the black hole (0.1$-$5~pc) and that the mean number of clouds along
any line of sight has to be limited ($\lesssim$6). Those criteria
reflect the necessity to have warm, optically thin emission at mid-IR
wavelengths and little cold dust. Note that the original CLUMPY models
included a mistake \citep{2010ApJ...723.1827N} which affected the
relative scaling of the contributions of the direct AGN emission
compared to the torus emission. Since we are comparing with the torus
emission alone, our results are not affected.

Using a redshift of $z=0.14$ and the following cosmological
parameters, H$_0$=71, $\Omega_M$=0.27, we find a luminosity distance
of 654~Mpc. We derive a silicate dust mass of $\sim$1\,300
M$_{\odot}$, for the 260~K component, assuming optically thin
emission, using the optical properties from \citet[][ Set 2 in their
Table~2 ]{1992A&A...261..567O}. For a gas-to-dust mass ratio of 100
this warm dust translates to a gas mass of
$\sim$1-2\,10$^{5}$~M$_{\odot}$. This should be considered a lower
limit to the mass of the warm medium that gives rise to the silicate
emission since the actual gas-to-dust mass ratio could be much higher,
in particular if dust is being destroyed by the harsh radiation field
originating from the accretion disk onto the black hole. However dust
residing inside clouds may be effectively shielded from destruction
\citep{2002ApJ...567L.107E}.

The shape of the SED, in particular the absence of MIPS 70, PACS 100
and 160~$\mu$m detections, constrains the amount of dust with
temperatures below 250~K that can be present. We find upper limits
(3$\sigma$) of 4, 3 and 4~mJy at 70, 100 and 160~$\mu$m (61, 96 and
140~$\mu$m rest wavelength), respectively. These limits imply that the
total amount of silicate dust with temperatures between 100 and 250~K
cannot exceed 10$^5$~M$_{\odot}$. The constraints set by the far-IR
upper limits on the presence of cooler dust are not as stringent; we
find that the maximum mass of silicate grains with temperatures
between $\sim$30 and 100~K cannot exceed 4\,10$^6$~M$_{\odot}$.

\section{Discussion}
\label{sec:discussion:hostless}
Silicate emission is diagnostic of the medium around the central
accretion region of active galactic nuclei (AGN). The presence of
silicate emission and the deduced grain temperatures have been
interpreted either in terms of a clumpy medium in the torus around the
black hole or in terms of clouds in the narrow line region
\citep{2005AN....326R.556S}. Both of these scenarios have been
proposed to explain the relatively low temperature of the silicate
grains derived from the IR feature ratio ($<$300~K). These scenarios
provide a way to have the silicate emission arise at a larger distance
from the accreting black hole than the inner disk. Dust present at the
inner wall of the accretion torus would be at much higher temperatures
than is observed. The sublimation temperature of silicates is
$>$1\,000~K \citep[e.g.][]{1994ApJ...421..615P}, consistent with the
hot dust 4-$\mu$m peak in the observed SED of \sage05.

Within a sample of $\sim$200 AGN and Ultra Luminous InfraRed Galaxies,
\citet{2007ApJ...655L..77H} find the strength of the 10~$\mu$m
silicate emission feature relative to the continuum to range between 1
and 1.5 for quasars. There is one outlier: \object{PG 1351+640} with a
ratio index of 2.1. \sage05, on the other hand, has a ratio of 3.6.
Thus we conclude that \sage05 is an extreme case of a population of
quasars with prominent silicate emission. The strong emission feature
is due to optically thin emission of relatively warm silicate grains.
The feature-to-continuum ratio is a function of the relative amount of
silicates that contribute to the optically thin emission with respect
to other components and the temperature of this material. Other
components here can refer to non-silicate grains, e.g. carbonaceous
grains, if present, that will give rise to a smooth continuum,
material in the optically thick accretion disk, but also to cooler
silicate grains, which will tend to diminish the contrast with the
continuum, if they cause self-absorption at 10~$\mu$m. Therefore, the
exceptional strength of the 10~$\mu$m emission feature may be in part
the result of a lack of cooler silicate grains, which cause
self-absorption in other sources.

\sage05 also stands out in the lack of far-IR emission. The source
remains undetected at 70, 100 and 160~$\mu$m (see Fig.~\ref{fig:maps})
and the IRS spectrum is dropping at wavelengths longer than the 18
$\mu$m Si-O bending resonance. Almost all of the mid-IR spectra of
quasars presented by \citet{2007ApJ...666..806N} show a flat or even
rising far-IR slope (in F$_{\nu}$, see also Fig.~\ref{fig:fig3}). The
only noticeably exception is \object{PG 1626+554}, which is the only
source in their study that does not require any cold dust component to
fit the IRS spectrum. We compare the IRS spectra of \sage05,
\object{PG 1351+640} and \object{PG 1626+554} in Fig.~\ref{fig:fig3}.
This comparison shows the exceptional energy distribution of \sage05.
The mid-IR spectrum flux level is comparable to the other sources but
those at other wavelengths are much lower than the comparison sources,
which are by themselves already very extreme sources. We are not aware
of any observationally based SED templates
\citep[e.g.][]{1994ApJS...95....1E,2007ApJ...666..806N,2008ApJ...676..286A}
of AGN+galaxy that do not over-predict the 70, 100 and 160~$\mu$m
fluxes when scaled to the mid-IR of \sage05. In those templates the
emission other than the mid-IR is dominated by contributions from the
host galaxy, either in the form of stellar photospheric emission
(optical) or cooler dust (far-IR).

Thus the IR emission of \sage05 is fully due to emission related to
the accretion onto the central black hole and the associated torus.
The two sources we chose for comparison (\object{PG 1351+640} and
\object{PG 1626+554}) are not typical sources but already represent
the previously known extreme cases in terms of feature-to-continuum
and lack of cold dust. There is a puzzling discrepancy between the
Infrared Astronomy Satellite (IRAS) photometry of \object{PG 1626+554}
and its IRS spectrum. The IRAS photometry is systematically higher
than would be expected based on the IRS spectrum (see
Fig.~\ref{fig:fig3}) which could be indicative of associated IR
emission which has been missed by the IRS spectroscopy.

Recently, \citet{2011arXiv1102.1425M} subtracted the host-galaxy
contribution from the observed 6-100~$\mu$m IR SEDs of a sample of
local galaxies hosting an AGN to derive the IR emission intrinsic to
the AGN. When we compare the observed SED of \sage05 with the
intrinsic SEDs presented in \citet{2010MNRAS.401..995M} we find a very
good agreement. This reinforces our interpretation that the IR regime
of this source is dominated by AGN activity.

We conclude that \sage05 is a source of which the emission is
dominated by contributions from the inner 10~pc. The SED puts strong
limits on the host galaxy emission.

Deep optical images are instrumental in determining the nature and
presence of the host galaxy. If there is a host galaxy, it is most
likely a red elliptical galaxy, since there are no signs of star
formation. Indeed, there seems to be some extended emission in the
image of the ESO-red\footnote{{from \tt
    http://www-wfau.roe.ac.uk/sss}} plate. The source is marginally
resolved. In the corresponding catalogue$^4$ the source is classified
as having little resemblance to a stellar source (point-source).
Comparing the full-width-at-half-max (FWHM) of the source to some
stellar sources in the same field we find that \sage05 is elongated
and its FWHM along the elongation is $\sim$ 1.4 times a typical
stellar FWHM. In Fig.~\ref{fig:SED} we show the SED template of an
elliptical galaxy (template Ell2) taken from {\it The SWIRE Template
  Library}\footnote{from \tt
  http://www.iasf-milano.inaf.it/$\sim$polletta/\\templates/swire\_templates.html}.
We scaled the template to avoid surpassing the observed photometry in
the near-IR, which is probably an over-estimation of the actual
contribution We conclude that the possible contribution of a host
elliptical to the total luminosity is small ($<$30\%).

\subsection{The origin of the PAH bands}
\begin{figure}[!t]
  \includegraphics[width=8.8cm]{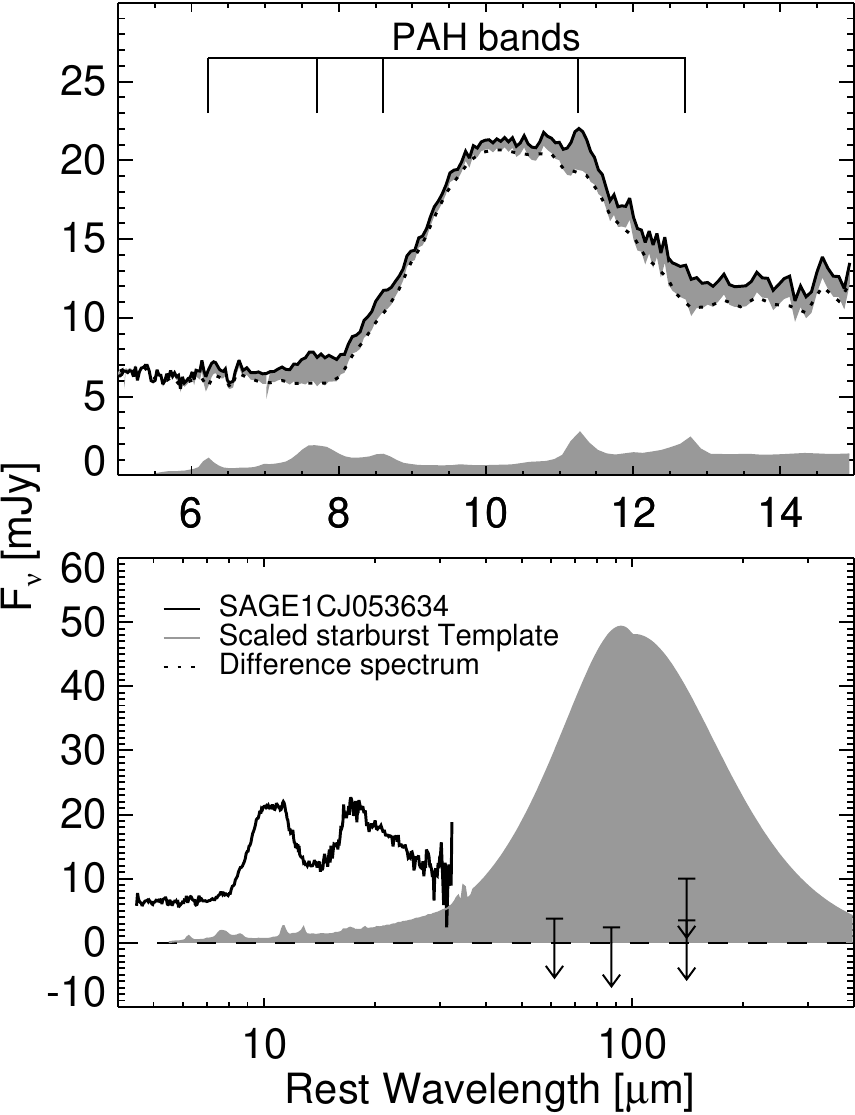}
  \caption{Decomposition of the mid-IR spectrum. The top panel shows
    the observed \sage05 spectrum in black. It is dominated by the
    silicate Si-O stretching feature. The position of the PAH features
    are indicated with tick-marks. The grey filled curves shows the
    spectrum of a starburst galaxy template from
    \citet{2011arXiv1102.1425M} scaled to the intensity of the PAH
    bands. We show this template twice, once alone (bottom) and once
    perched on top of the silicate 10~$\mu$m feature. The bottom panel
    shows the full IR SED of \sage05, including the far-IR photometric
    3~$\sigma$ upper limits, compared to the scaled template. The
    starburst template clearly over-predicts the far-IR flux levels. }
  \label{fig:starburst}
\end{figure}
The presence of PAH emission bands, albeit faint, may originate from
star formation in the ISM of the host. This is an assumption which is
often made when decomposing the IR SED into an AGN and a star
formation component.

We have investigated this possibility by comparing the observed PAH
emission with the starburst galaxy templates of
\citet{2011arXiv1102.1425M}. These templates were derived by mean
averaging similar SEDs from groups of galaxies that cover the full
range of IR colours of starburst and Luminous InfraRed galaxies in the
local Universe. The template spectrum agrees quite well with the
observed PAHs spectrum of \sage05 (see Fig.~\ref{fig:starburst}).
However, the starburst template over-predicts the upper limits of the
far-IR emission by more than a factor of 5. We conclude that the faint
PAH emission of \sage05 is not linked directly to star formation.

An alternative explanation could be that the PAH features arise from
the host elliptical galaxy \citep[e.g.][]{2005ApJ...632L..83K}.
Several local ellipticals exhibit PAH features with exceptional band
ratios, the 11.3~$\mu$m band being strong compared to the other bands,
which has been interpreted as these PAH features originating in a post
starburst phase from matter condensed in shells around evolved carbon
stars \citep{2010ApJ...721.1090V}. This possibility is interesting
since it would naturally explain the lack of far-IR emission. Is
should be stressed that those local ellipticals which exhibit strong
PAH features have active star formation and the corresponding strong
far-IR emission.

Because the PAH features in the IRS spectrum of \sage05 are weak and
perched on top of very strong silicate emission features, it is not
possible to measure reliably the band ratios. We have further
investigated the possible origins of the PAHs from a post starburst
in the host red elliptical by comparing with \object{NGC 708}. This
source exhibits the strong 11.3~$\mu$m feature on a weak continuum
\citep{2008ApJ...684..270K}. We have scaled the observed mid-IR IRS
spectrum to the overall IRAS photometry. This assumes that the nuclear
IRS spectrum is representative of the entire host. We have
subsequently scaled the overall NGC 708 SED to not surpass the optical
photometry of \sage05. In this way we obtain a maximum template SED of
the possible host galaxy. We find that in this scaled template SED the
strength of the 11.3~$\mu$m band is $\sim$50 \% of that observed in
\sage05. We conclude that the PAHs we observe in \sage05 may have an
origin in the host galaxy, without being directly linked to star
formation.

\section{\sage05 as a sub-class: IR selection criteria}
\label{sec:ir-criterion}
\begin{figure}[!t]
  \includegraphics[width=8.8cm]{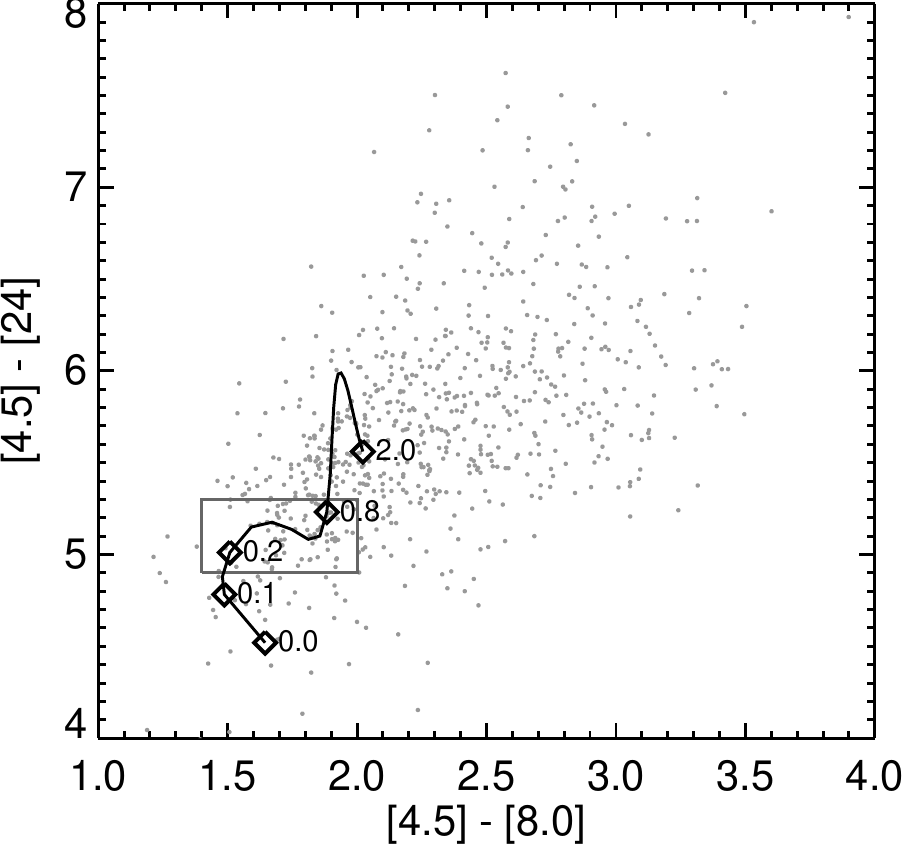}
  \caption{[8.0]--[4.5] vs [24]--[4.5] colour-colour diagram based on
    detections off the main part of the LMC (grey dots). The grey box
    encompasses the region in which sources with similar magnitudes
    and colours as \sage05 are located. The black line shows the
    simulated colours of a source like \sage05 as a function of
    redshift. }
  \label{fig:CMD}
\end{figure}
It is interesting to try to estimate how rare objects like \sage05
are. We use the photometric catalogue of SAGE-LMC to search for
sources with similar SEDs. We focus on the sources observed off the
main body of the LMC \citep{2008AJ....136...18W}. This choice is to
avoid confusion with sources in the LMC, i.e. to ensure that the
sources with IR excess are background galaxies. The total number of
sources with [4.5]--[8.0] colour redder than stars ($>$1) is $\sim$
1\,100. Of these, 735 are also detected at 24~$\mu$m and are shown in
Fig.~\ref{fig:CMD} as grey dots. We have constructed a smooth template
SED based on the existing photometry and the IRS spectrum. The
smoothed version has been obtained by interpolating the photometry on
a finer wavelength grid and extrapolating the SED through the
3$\sigma$ upper limits at far-IR wavelengths. We simulate the
photometry as a function of redshift by convolving the template with
the filter profiles. We show the synthetic colours in
Fig.~\ref{fig:CMD}, where the {\it Spitzer} [4.5]--[8.0] and
[4.5]--[24] colours best trace the characteristic shape of the SED.
However these colours are still not very efficient in detecting the
most striking features of this particular quasar, i.e. the strong
silicate 10~$\mu$m emission and the lack of cold dust. Taking into
account also the magnitudes and the near IR colours we find 69 sources
with colours similar to \sage05. All of these sources are located in
the grey box in Fig.~\ref{fig:CMD}. There are no sources matching this
SED at low redshifts ($z<$0.15), i.e. very bright. Beyond
$z$$\approx$0.8 this kind of source would be too faint to be detected
in all of the bands in SAGE-LMC but could be detected in deep-field
surveys. Many, if not most, of these sources may be more normal AGN
hosting galaxies. Indeed \citet{2009ApJ...701..508K} find $\sim$5\,000
candidate AGN sources behind the LMC in the SAGE-LMC catalogue. These
authors use primarily the IRAC colours, which trace the hot emission
from the accretion disk and \sage05 is included in their candidate
list, albeit with somewhat atypical colours. We conclude that sources
like \sage05 are relatively rare in the local universe ($z<$0.15) but
that there is a possible population farther away. This population
could be as large as $\sim$10\% (69/735) of the IRAC detected sources.
Taking into account the fact that mostly type I sources display the
silicate band in emission and that at these redshifts there are
similar numbers of type I and type II AGNs, this fraction further
decreases to $\sim$5\%.

The fact that sources like \sage05 are perhaps less rare in the
distant universe than at low redshift may reflect the diminishing star
formation activity trend with the age of the Universe. If \sage05
represents a galaxy at the end of its AGN activity when the star
formation has already ceased -- as indicated by its very low far-IR
luminosity -- then such galaxies are expected to be more common in the
more distant Universe, up to redshifts approaching the typical
formation epoch of massive early-type galaxies around $z\sim2$--3.
However, IR spectroscopy is essential for discriminating between this
kind of extreme silicate emitters and more regular active galaxies.

\begin{figure}[!t]
  \includegraphics[width=8.8cm]{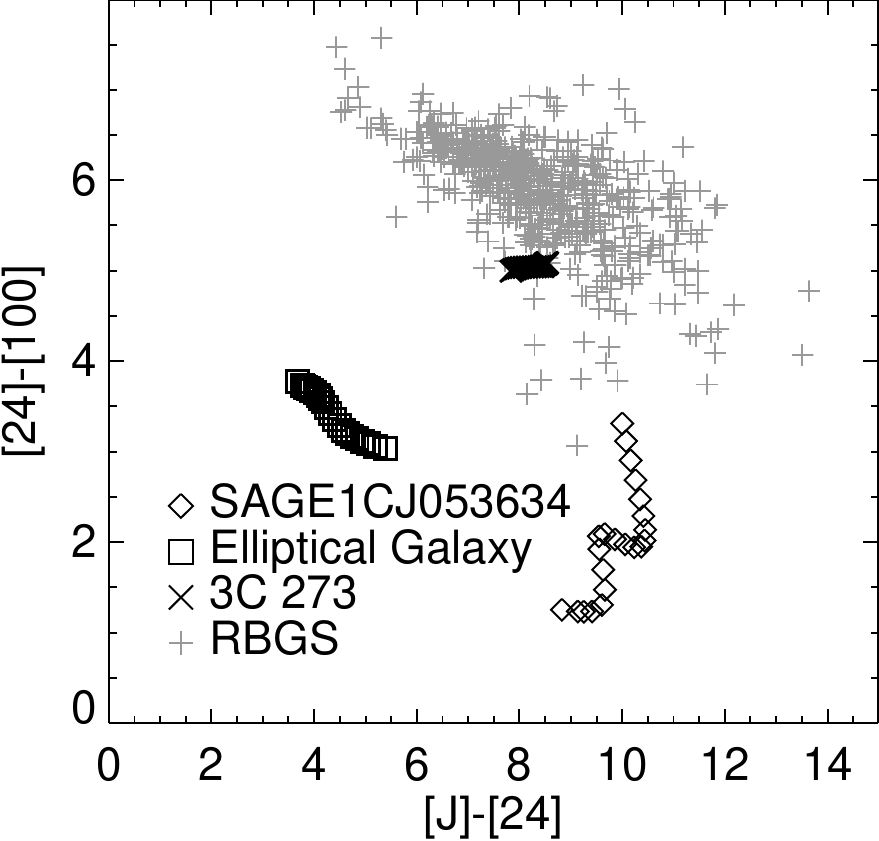}
  \caption{2MASS [J]-- MIPS [24] vs MIPS [24]-- PACS [100]
    colour-colour diagram based on detections off the main part of the
    LMC (grey dots). The black diamonds show the simulated colours of
    a source like \sage05 as a function of redshift. We also show for
    the location of an elliptical galaxy and a more typical strong AGN
    source (3C 273) in this diagram. The redshift increases from 0 to
    2 from left to right in the diagram for each source. The crosses
    show the approximate location of bright galaxies, using IRAS
    photometry based on the IRAS Revised Bright Galaxy Sample \citep[
    RGBS,][]{2003AJ....126.1607S}. }
  \label{fig:CMD2}
\end{figure}

We also present a diagram which may be useful to discover similar
objects using the many {\it Herschel} surveys that are ongoing
(Fig.~\ref{fig:CMD2}). Since the extreme AGN activity is most
prominently visible in the mid-IR, the near-IR to mid-IR colour of
such sources is very red while the mid-IR to far-IR colour is
extremely blue. Figure~\ref{fig:CMD2} clearly shows that selecting
sources with a [MIPS 24]--[PACS 100] colour below 3 would be very
efficient in selecting further candidate objects.

\section{Conclusions}
\label{sec:conclusion}
We have presented the SED and SAGE-Spec {\it Spitzer}/IRS spectrum
from 5$-$37~$\mu$m of \sage05, an IR point-source detected by {\it
  Spitzer}/SAGE-LMC. The spectrum shows a redshifted ($z$=0.14)
silicate emission feature with an exceptional feature-to-continuum
ratio and weak PAH emission bands, which may arise from the host
galaxy. We have constructed the energy distribution from the UV to
radio wavelength. The energy distribution is peculiar because it lacks
a clear stellar counterpart and far-IR emission. We find that the
entire IR energy distribution and the IR spectrum can be understood as
emission originating from the inner active region around an accreting
black hole. The hot dust near the accretion disk gives rise to the
optically thick continuum, which peaks at 4~$\mu$m and is roughly
constant in F$_\nu$ for the IRS wavelength range. The strong silicate
features, on the other hand, may arise from optically thin emission of
dusty clouds within $\sim$10~pc around the black hole. As such \sage05
can serve as a template of the emission from the AGN in other
galaxies. Extreme sources dominated by silicate emission such as
\sage05 are rare in the local universe, but are perhaps more common at
earlier epochs. We have presented colour-colour criteria as tools to
search for candidate sources like those based on {\it Spitzer} as well
as for {\it Herschel} data.

\begin{acknowledgements}
  This research has made use of NASA's Astrophysics Data System
  Bibliographic Services, the VizieR catalogue access tool, CDS,
  Strasbourg, France, the NASA/ IPAC Infrared Science Archive, which
  is operated by the Jet Propulsion Laboratory, California Institute
  of Technology, under contract with the National Aeronautics and
  Space Administration and the ROSAT Data Archive of the
  Max-Planck-Institut f\"{u}r extraterrestrische Physik (MPE) at
  Garching, Germany. SH would like to thank N.~Jetha and S.~Temporin
  for interesting and very instructive discussions about active
  galaxies. The authors thank Neal Jackson and Ian Browne for careful
  reading of an earlier version of the manuscript.
\end{acknowledgements}
  
\bibliographystyle{aa}
\bibliography{hony}
  
\end{document}